\documentclass[aps,prl,twocolumn,amsmath,amssymb,superscriptaddress]{revtex4-1}
\usepackage{epsfig,amssymb}
\usepackage{graphicx}
\usepackage{dcolumn}
\usepackage{bm}
\usepackage{dcolumn}
\usepackage{multirow}
\begin{document}

\title{Generation of plasmonic hot carriers from d-bands in metallic nanoparticles}

\author{Lara Rom\'an Castellanos}
\affiliation{Department of Materials, Imperial College London, London SW7 2AZ}
\affiliation{Department of Physics, Imperial College London, London SW7 2AZ}
\author{Juhan Matthias Kahk}
\affiliation{Department of Materials, Imperial College London, London SW7 2AZ}
\author{Ortwin Hess}
\affiliation{Department of Physics, Imperial College London, London SW7 2AZ}
\author{Johannes Lischner}
\affiliation{Department of Physics, Imperial College London, London SW7 2AZ}
\affiliation{Departments of Materials, Imperial College London, London SW7 2AZ}

%\begin{abstract}
%We present an approach to master the well-known challenge of %calculating the contribution of d-bands to plasmon-induced hot %carrier rates in spherical metallic nanoparticles. Specifically, we %first demonstrate how the widely used spherical well model for the %nanoparticle wavefunctions can be derived from atomistic %first-principles methods using the envelope function technique. We %then generalize this approach to flat d-bands. Using Fermi's golden %rule, we calculate the generation rates of hot carriers due to %transitions from either d-sptate to sp-sptate transitions or sp-sptate to %sp-sptate transitions. We apply this formalism to spherical silver %nanoparticles and find that for small nanoparticles with radii less %than 60 nanometers sp-sptate to sp-sptate transitions dominate hot %carrier rates. For larger nanoparticles, d-sptate to sp-sptate %transitions give the larger contribution.
%\end{abstract}

%LRC: Maybe cite and fully read: Berglund, C. N. & Spicer, W. E. Photoemission studies of copper and silver: experiment. Phys. Rev. 136, A1044–A1064 (1964).

\begin{abstract}
We present an approach to master the well-known challenge of calculating the contribution of d-bands to plasmon-induced hot carrier rates in metallic nanoparticles. We generalise the widely used spherical well model for the nanoparticle wavefunctions to flat d-bands using the envelope function technique. Using Fermi's golden rule, we calculate the generation rates of hot carriers after the decay of the plasmon due to transitions from either a d-band state to an sp-band state or from an sp-band state to another sp-band state. We apply this formalism to spherical silver nanoparticles with radii up to 20~nm and also study the dependence of hot carrier rates on the energy of the d-bands. We find that for nanoparticles with a radius less than 2.5~nm sp-band state to sp-band state transitions dominate hot carrier production while d-band state to sp-band state transitions give the largest contribution for larger nanoparticles.
\end{abstract}

\maketitle

\section{Introduction}

%LRC: Added sentence about other decay channels:
There is currently significant interest in the properties of plasmon-induced hot carriers in metallic nanostructures. Such nanostructures absorb sunlight by generating localized surface plasmons (LSPs) which can decay into electron-hole pairs via the Landau damping mechanism. Other damping mechanisms exist, such as defect- or phonon-mediated processes, but Landau damping is the dominant mechanism in small nanoparticles~\cite{Khurgin2019}. The resulting energetic carriers can be harnessed in new applications for catalysis \cite{Halas2011,Baffou2014} or solar energy conversion~\cite{Govorov2013,Li2016,Narang2016,Hartland2017}. %However, Landau damping dominates for nanoparticles smaller than the mean free path and it is also the most efficient mechanism for the generation of hot carriers.
%In the uptaded manuscript, we mention that other loss mechanisms are present in real nanoparticles but Landau damping dominates for the small nanoparticles under consideration. We have added the citation to the suggested paper as a reference that support this idea.

%LRC: When we say that in Au and Cu there are e- excited from d-s above the Fermi, do we mean in Ag that does not happen? In Cu you need to include interband to predict correct optical properties, see: Cu Nanoshells: Effects of Interband Transitions on the Nanoparticle Plasmon Resonance Wang 2005

%LRC: I have added sth about hot holes

Most experiments employ traditional plasmonic metals, such as Ag or Au, as these materials exhibit strong plasmonic resonances in their absorption spectrum. The electronic structure of these materials is characterized by a dispersive band of mixed s- and p-state character (referred to as the sp-band) which crosses the Fermi level and multiple occupied d-bands with a comparably flat dispersion~\cite{Christensen1972,Sundararaman2014}. If the d-bands are sufficiently close to the Fermi energy, it is possible to excite electrons from the d-bands into the sp-band~\cite{DuChene2018,Berdakin2020}. For example, Barman and coworkers~\cite{Barman2015} measured a photocurrent arising from hot d-band holes in gold nanoparticles, but the relative importance of such d-to-sp transitions compared to transitions between sp-band states in nanoparticles has not yet been studied in detail. 

%While the d-bands are quite far from the Fermi level in Ag, they are much closer in the cases of Au and Cu and it is possible to excite electrons from the d-bands into sp-band states above the Fermi level~\cite{DuChene2018}. For example, Barman et al.~\cite{Barman2015} measured a photocurrent generaterated by the hot holes resulting from the d to sp inter-band transition of electrons in gold nanoparticles. However, it is not clear how important this process is compared to excitations within the sp-band. 

%LRC: I am adding another model for interband, only for optical properties in Cu.
%Other authors have included the effect of interband transitions on the optical properties of copper using Mie theory~\cite{Wang2005}. 

Theoretical modelling of hot electron processes allows valuable insights into experimental observations. To describe the electronic structure of noble metal nanoparticles, several groups have solved the Schr\"odinger equation of electrons in a spherical well~\cite{Manjavacas2014,DalForno2018,Ekardt1984,Crai2017,Yannouleas1993,Kumarasinghe2015,Martins1981}. While this approach allows the description of experimentally relevant nanoparticles with radii of the order of several tens of nanometers, it does not capture the contribution of d-bands. On the other hand, first-principles methods, such as density-functional theory (DFT), allow the description of d-bands~\cite{Bernardi2015,Sundararaman2014,Kuisma2015,Rossi2017b}, but can only be applied to very small nanoparticles (typically only a few hundred atoms)~\cite{Ma2015}.

In this paper, we present an approach that bridges atomistic and continuum electronic structure theories of metallic nanoparticles via the envelope function technique~\cite{Luttinger1955} and allows the description of nanoparticle states derived from d-bands in metallic nanoparticles with large radii. We apply this approach to silver nanoparticles and show that sp-band to sp-band transitions give the dominant contribution to hot carrier rates for nanoparticles with radii less than approximately 2.5~nm. However, as the rate of d-band state to sp-band state transitions increases more quickly with the nanoparticle radius than the rate of sp-band state to sp-band state transitions, we predict that d-to-sp transitions dominate for larger nanoparticles.

\section{Methods}

\subsection{Hot carrier generation rates}
We employ the framework of Dal Forno, Ranno and Lischner to calculate hot carrier generation rates in spherical nanoparticles \cite{DalForno2018}. In this
approach, the total number of plasmon-induced hot electrons generated per unit time in a nanoparticle illuminated by light (polarized along the z-direction) of frequency $\omega$ is determined using Fermi's golden rule according to
\begin{equation}
  N(\omega) = \frac{4\pi}{\hbar} \sum_{if}
  |M_{if}|^2 \delta(\hbar \omega - [E_f-E_i]),
\label{eq:Fermi}
\end{equation}
where $M_{if} = \langle \psi_f | \Phi_{pl}(\omega) | \psi_i \rangle$ denotes
the matrix element of the total potential $\Phi_{pl}$ which is often
calculated using the quasistatic approximation. In this approximation, the potential inside the spherical nanoparticle is $\Phi_{pl}(\omega)=-e E_{0}\frac{\epsilon(\omega)-1}{\epsilon(\omega)+2}z$, where $e$ denotes the electron charge, $E_0$ is the strength of the externally applied electric field and $\epsilon(\omega)$ denotes the bulk dielectric function of the material. For the latter, we use a Drude model
\begin{equation}
    \epsilon(\omega) = \epsilon_b - \frac{\omega^2_0}{\omega^2-i\omega \gamma_P},
\end{equation}
where $\gamma_P$ denotes the plasmon linewidth, $\omega_0$ is the bulk plasmon frequency and $\epsilon_b$ is the dielectric constant due to the background screening by the polarizable d-bands. Also, $\psi_i$ and $\psi_f$ denote quasiparticle wavefunctions of occupied and empty states with energies $E_i$ and $E_f$, respectively. Note that we have neglected effects arising from finite quasiparticle lifetimes in Eq.~\eqref{eq:Fermi} which give rise to anti-resonant contributions~\cite{DalForno2018}. Such transitions only give rise to low-energy carriers and therefore do not play an important role in applications, such as photodetection and photocatalysis. A factor of two arising from spin is taken into account.

\subsection{Envelope function method}
We calculate the electronic wavefunctions of the nanoparticle using the envelope function method originally developed by Kohn and Luttinger to describe the electronic structure of charged defects in semiconductors~\cite{Luttinger1955,Kittel1976}. In this approach, one assumes that the electronic structure of the defect-free material is
known, i.e. that the eigenstates $\phi_{n\mathbf{k}}$ and eigenvalues $\epsilon_{n\mathbf{k}}$ (with $n$ and $\mathbf{k}$ denoting the band index and the crystal momentum, respectively) of the crystal Hamiltonian
$H_{\text{crys}}$ have been determined by solving the Schr\"odinger equation
\begin{equation}
  H_{\text{crys}} \phi_{n\mathbf{k}}(\mathbf{r}) = \epsilon_{n\mathbf{k}} \phi_{n\mathbf{k}}(\mathbf{r}).
\end{equation}
Nowadays, this task can be carried out routinely using first-principles
methods, such as density-functional theory or the more advanced GW approach. 

Next, a perturbation $\delta V$ is considered which breaks the
discrete translational invariance of the infinite crystal. For charged defects, the perturbation is the screened Coulomb potential induced by the defect. For a nanoparticle, this is the spherical well potential which
confines the electrons inside the nanoparticle. The eigenstates $\psi$ (with corresponding eigenenergies $E$) of the perturbed Hamiltonian
$H_{\text{crys}} + \delta V$ can be constructed as linear combinations of the
crystal states according to
\begin{equation}
  \psi(\mathbf{r}) = \sum_{n\mathbf{k}} c_{n\mathbf{k}} \phi_{n\mathbf{k}}(\mathbf{r}),
  \label{eq:psi}
\end{equation}
with $c_{n\mathbf{k}}$ being complex coefficients. Inserting this ansatz into the Schr\"odinger equation and multiplying from the right with $\phi^*_{n'\mathbf{k'}}$ results in a matrix equation for the coefficients. Assuming that the perturbation does not mix different bands leads to 
\begin{equation}
  \epsilon_{n\mathbf{k}} c_{n\mathbf{k}} + \sum_{\mathbf{k'}} \langle
  \phi_{n\mathbf{k}} | \delta V | \phi_{n\mathbf{k'}} \rangle c_{n\mathbf{k'}} = E c_{n\mathbf{k}}.
\end{equation}

Inserting the Fourier transform $c_{n\mathbf{k}} = \int d^3r
c_n(\mathbf{r}) \exp(-i\mathbf{k} \cdot \mathbf{r})$ yields
\begin{equation}
  \label{eq:envelope}
  \epsilon_{n\mathbf{p} } c_{n}(\mathbf{r})
  + \delta V(\mathbf{r}) c_{n}(\mathbf{r}) = E c_{n}(\mathbf{r}),
\end{equation}
with $\mathbf{p}=-i\hbar\nabla$ denoting the momentum operator.

For a band with a parabolic dispersion, i.e. $\epsilon_{n\mathbf{k}} = \hbar^2\mathbf{k}^2/(2m^*)$ (with $m^*$ denoting the effective mass of the band), the equation for the envelope function $c_{n}(\mathbf{r})$ has the same form as the Schr\"odinger equation for an electron in the potential $\delta V$. This demonstrates how the spherical well approximation for a nanoparticle can be derived from an atomistic model.

Often, an additional approximation is invoked for the wavefunctions~\cite{Kittel1963}. Specifically, it is often found that the linear combination in Eq.~\eqref{eq:psi} is dominated by states near a specific crystal momentum $\mathbf{k}_0$. For semiconductors, this is typically the crystal momentum corresponding to the valence and conduction band extrema. In this case, the integral over $\mathbf{k}$ can be carried out analytically and one finds
\begin{equation}
    \psi_n(\mathbf{r}) = c_n(\mathbf{r}) u_{n\mathbf{k}_0}(\mathbf{r}),
\end{equation}
where $u_{n\mathbf{k_0}}(\mathbf{r})$ is a lattice-periodic function. Assuming that $c_n(\mathbf{r})$ is normalized, normalization of $\psi_n$ requires that 
\begin{equation}
    \int_{V_{uc}} d^3r |u_{n\mathbf{k}_0}|^2 = V_{uc}.
\end{equation}
For metals, it is much less clear that this simplified form for $\psi$ is justified and which crystal momentum $\mathbf{k}_0$ should be chosen. However, this form is highly advantageous for analytical treatment and will be adopted in this work. Future work will aim to assess the accuracy of this approximation.

\subsection{Electronic states of spherical nanoparticles}
For a parabolic band and a perturbation with spherical symmetry, we can express the kinetic energy operator in Eq.~\eqref{eq:envelope} in spherical coordinates and use a separation of variable ansatz for $c_n(r,\theta,\phi)$. The angular part of the envelope function is described by the spherical harmonics $Y_{lm}$ (where $l$ and $m$ denote the orbital and magnetic quantum numbers, respectively) and the radial part $R_{\nu l}$ (with $\nu$ denoting the principal quantum number) is the solution of
\begin{equation}
\label{eq:radial}
  \frac{d}{dr} \left( r^2 \frac{dR_{\nu l}}{dr} \right) +
  \frac{2m^*r^2}{\hbar^2} \left( E_{\nu l} - \delta V(r) \right) R_{\nu l} -
  l(l+1) R_{\nu l}
  = 0.
\end{equation}

Analytic solutions of Eq.~\eqref{eq:radial} exist for the \emph{infinite} square well potential, i.e. a potential that is zero inside the nanoparticle and infinite outside. In this case, the radial solutions are proportional to the spherical Bessel functions of the first kind, i.e. $R_{\nu l}(r) \propto j_l(k_{\nu l}r)$ with corresponding eigenenergies $E_{\nu l}=\hbar^2k^2_{\nu l}/(2m^*)$. The allowed values of $k_{\nu l}$ are determined by the boundary condition $R_{\nu l}(r_0)=0$ with $r_0$
denoting the radius of the nanoparticle. For large nanoparticles, we can approximate $j_l(x)$ by $\sin(x-l\pi/2)/x$ and find
\begin{equation}
  E_{\nu l} = \frac{\hbar^2 \pi^2}{2m^* r^2_0} \left(  \nu + l/2 \right)^2,
\end{equation}
with $\nu \geq 1$ and $l \geq 0$ being integers.

For later use, we also calculate the corresponding angular momentum resolved density of states, $g_l(\epsilon) = \sum_\nu \delta(\epsilon-E_{\nu l})$ for a parabolic sp-band and find
\begin{equation}
  g^{sp}_{l}(\epsilon) = \sqrt{ \frac{m^* r^2_0}{2\hbar^2\pi^2
      \epsilon}} \Theta(\epsilon - E^\text{min}_l)
\end{equation}
with $E^{\text{min}}_{l}=\hbar^2\pi^2(l+2)^2/(8m^*_n r^2_0)$ denoting the
smallest energy for a fixed value of $l$. Here, we have assumed that
the spacing between energy levels is sufficiently small that the sum
over $\nu$ can be replaced by an integral. 

\subsection{D-band states in spherical nanoparticles}
The envelope function method is not limited to parabolic dispersion relations. For any dispersion relation
$\epsilon_{n\mathbf{k}}$ a corresponding equation for the envelope function can be derived by replacing $\mathbf{k}$ by $-i\hbar \nabla$, see Eq.~\eqref{eq:envelope}. In practice, however, it is difficult to carry out this procedure because the dependence of the band structure on crystal momentum is not known analytically. As a first step towards understanding the role of d-bands in hot carrier generation, we invoke a drastic approximation and consider the limit of a perfectly flat band, i.e. $\epsilon_{n\mathbf{k}} = \epsilon_d$. For this dispersion, it is straightforward to solve the envelope function equation for an infinite square well. In fact, the solutions of the parabolic case are also solutions of the flat band case, but all have the same eigenvalue $\epsilon_d$.

As all d-band states have the same energy and there are in principle infinitely many solutions to the spherical well equation, some regularization procedure is required to obtain meaningful results. This is achieved by imposing that the envelope
function should not oscillate faster in the radial direction than the spacing between atoms. This leads to the condition that only solutions with quantum numbers $\nu < \nu_\text{max}(l) =\left( \frac{r_{0}}{a_0} - \frac{l}{2} \right)$ are allowed where and $a_0$ is the lattice constant of the crystal. The maximum allowed value of $l$ is then determined by the condition that $\nu \geq 0$, i.e. $l_{\text{max}}=2  r_0/a_0$. Note that our final results for the hot carrier rates do not depend on $a_0$. No regularization procedure is needed for a parabolic band as envelope functions with unphysical fast oscillations have very high energies and do not influence our results. 

%JL: find a place for this
%In the following, we will restrict ourselves to the case of $N_{d}=5$ degenerate occupied flat bands and a single parabolic band that crosses the Fermi level. 

\subsection{Matrix elements}
Having determined the quasiparticle energies and wavefunctions, we are now in a position to evaluate Eq.~\eqref{eq:Fermi}. We replace the summations over initial and final states by sums over the quantum numbers $\nu$, $l$ and $m$ for each occupied and empty band. Using Eq.~\eqref{eq:envelope} for the nanoparticle wavefunctions, the 
matrix elements are given by 
\begin{align}
  M_{if} &= -eE(\omega) \int d^3r \psi^*_f(\mathbf{r}) z \psi_i(\mathbf{r}) \\
         &=
  -eE(\omega) \int d^3r c^*_{n\nu lm}(\mathbf{r})u^*_{n\mathbf{k}_0}(\mathbf{r}) z
  c_{n'\nu' l'm'}(\mathbf{r}) u_{n'\mathbf{k}_0}(\mathbf{r})
\end{align}
with $E(\omega)=E_0 (\epsilon(\omega)-1)/(\epsilon(\omega)+2)$.

Splitting the integral into a sum over all unit cells in the nanoparticle and integrals over each unit cell yields
\begin{equation}
  \label{eq:Mif}
  M_{if} = -eE(\omega) \left( z_{n\nu\nu' ll'mm'}^\text{env} \delta_{nn'} +
    z^\text{crys}_{nn'}  \delta_{\nu\nu'} \delta_{ll'} \delta_{mm'} \right),
\end{equation}
where the orthonormality of the $u_n$'s and the $c_n$'s was used. Here, $z^\text{crys}_{nn'}=\int' d^3r u^*_{n\mathbf{k}_0}(\mathbf{r})z u_{n'\mathbf{k}_0}(\mathbf{r})/V_{uc}$ (where the prime on the integral sign indicates integration over a unit cell with volume $V_{uc}$) denotes the transition
dipole moment of the vertical transition between the bands $n$ and $n'$ in the bulk material and 
\begin{align}
z^\text{env}_{n\nu \nu' ll' mm'} &= \int d^3r
c^*_{n\nu lm}(\mathbf{r}) z c_{n\nu' l'm'}(\mathbf{r}) \\ &= [A_{lm} \delta_{mm'}
  \delta_{l+1,l'} + B_{lm} \delta_{mm'} \delta_{l-1,l'}] z^\text{env}_{\nu
  l \nu' l'}.
\end{align}
Here, we defined
\begin{align}
A_{lm} &=\sqrt{\frac{(l+1-m)(l+1+m)}{(2l+1)(2l+3)}}, \\
B_{lm} &=\sqrt{\frac{(l-m)(l+m)}{(2l-1)(2l+1)}}
\end{align}
and also \cite{Weick2005,Yannouleas1985} 
\begin{equation}
z^\text{env}_{\nu
  l \nu' l'}=\int dr r^3
  R^*_{\nu l} R_{\nu' l'} = \frac{2\hbar^{2}}{m^{*}r_{0}} \frac{ \sqrt{\epsilon_{\nu l} \epsilon_{\nu' l'}}}{(\epsilon_{\nu l} - \epsilon_{\nu' l'})^{2}}.
  \label{eq:yannouleas}
\end{equation}
The last equality is valid for large nanoparticles~\cite{Weick2005,Yannouleas1992a}.

The structure of Eq.~\eqref{eq:Mif} is interesting: for transitions between nanoparticle states originating from the same band, the matrix element is determined entirely by the envelope functions and the well-known angular momentum selection rules are recovered. In contrast, only the microscopic transition dipole moment determines the strength of the matrix element for the inter-band transitions.

To evaluate $z^\text{crys}$, an expression for $u_{n\mathbf{k}_0}(\mathbf{r})$ is needed. In principle, this can be obtained straightforwardly from a first-principles band structure calculation, but it is not a priori clear how $\mathbf{k}_0$ should be chosen. To make progress, we choose $\mathbf{k}_0$ as the $\Gamma$-point of the first Brillouin zone, but interpret $u_{n\Gamma}$ as an effective function that represents the average behaviour of the band $n$ in the whole Brillouin zone. Using a tight-binding ansatz, we express the lattice periodic function describing the sp-band as
\begin{align}
    u_{sp,\Gamma}(\mathbf{r}) = \sqrt{V_{uc}} \sum_\mathbf{R} \left[ \alpha_p \phi_p(\mathbf{r}-\mathbf{R}) + \alpha_s \phi_s(\mathbf{r}-\mathbf{R})  \right],
\end{align}
where $\phi_s$ and $\phi_p$ denote \emph{atomic} s- and p-orbitals and $\mathbf{R}$ is a lattice vector. The coefficients $\alpha_p$ and $\alpha_s$ describe the relative contributions of s- and p-states and can be determined from a Mulliken analysis of the Bloch states obtained from a first-principles calculations as explained below. Note that the true $u_{sp,\Gamma}$ obtained from a band structure calculation would only have s-contributions by symmetry. 

%JL: added a bit here in response to reviewer 1
If a similar tight-binding ansatz is used for the d-band (but without any admixture of non-d states) and inter-atomic contributions to the dipole matrix element are neglected (assuming that the wavefunction overlap between neighbouring atoms is small which is consistent with the fundamental assumption of the tight-binding ansatz), we find
\begin{equation}
    z^\text{crys}_{d,sp} = \alpha_p z^\text{at}_{dp}
\end{equation}
with $z^\text{at}_{dp}$ denoting the dipole moment for an \emph{atomic} d-to-p transition which can be obtained from a first-principles calculation. Note that the atomic dipole selection rule forbids transitions from d-states to s-states. 

\subsection{Final expressions for the hot carrier rates in spherical nanoparticles}

The total number of hot electrons $N(\omega)$ can be written as a sum of $d$-band state to $sp$-band state transitions $N^{d \rightarrow sp}$ and $sp$-band state to $sp$-band state
transitions $N^{sp \rightarrow sp}$ (there are no transitions among d-band states as they are fully occupied). Using our results for the matrix element of d-to-sp transitions, we find 
\begin{widetext}
\begin{align}
    N^{d\rightarrow sp}(\omega) &= \frac{4\pi}{\hbar} N_{d}\left| eE(\omega)z^\text{crys}_{d,sp} \right|^2 \sum_{l=0}^{l_\text{max}} (2l+1) \sum_{\nu=1}^{\nu_\text{max}(l)} \delta(\epsilon_d + \hbar \omega - E_{\nu l}) \Theta(E_F-\epsilon_d)\Theta(E_{\nu l}-E_F) \\
   & = \frac{8\pi}{\hbar} \frac{N_{d}}{\sqrt{\tilde{\epsilon}(\epsilon_{d} +\hbar \omega)}}\left| eE(\omega)z^\text{crys}_{d,sp} \right|^2 \Theta(E_F -\epsilon_d) \Theta(\epsilon_d + \hbar \omega - E_F) \left[ \frac{\epsilon_d + \hbar\omega}{\tilde{\epsilon}} - \sqrt{\frac{\epsilon_d + \hbar\omega}{\tilde{\epsilon}}} + \frac{1}{4} \right],
   \label{eq:Nds}
\end{align}
\end{widetext}
where we defined $\tilde{\epsilon}=\hbar^2\pi^2/(2m^{*}r_{0}^{2})$ and the sum over $\nu$ has been transformed to an integral. Also, $E_F$ denotes the Fermi level and $N_d=5$ denotes the number of d-bands. For large nanoparticles, $(\epsilon_d + \hbar\omega)/\tilde{\epsilon} \gg 1$ and the last two terms in the brackets can be neglected resulting in
\begin{widetext}
\begin{align}
    N^{d\rightarrow sp}(\omega)  = \frac{8\pi}{\hbar} N_{d} \left( \frac{2m^*r_0^2}{\hbar^2\pi^2} \right)^{3/2} \left| eE(\omega)z^\text{crys}_{d,sp} \right|^2 \Theta(E_F -\epsilon_d) \Theta(\epsilon_d + \hbar \omega - E_F)  \sqrt{\epsilon_d + \hbar\omega}.
    \label{eq:Nds_large}
\end{align}
\end{widetext}
%JL: added this paragraph in response to reviewer 1
Note that the d-to-sp transition rate is proportional to $r_0^3$ and thus scales with the volume of the nanoparticle. This is expected as in the bulk material only d-to-sp transitions are possible (there are no vertical sp-to-sp transitions in the bulk). As discussed above, the matrix element of d-to-sp transitions does not depend on the nanoparticle size, but the number of available d-to-sp transitions does which leads to the calculated volume dependence. 

For the sp-to-sp transitions, we express the summations over initial and final states as integrals over the sp-band density of states according to
\begin{widetext}
\begin{align}
   N^{sp \rightarrow sp}(\omega) & = 8\pi\hbar^{3} \left| \frac{eE(\omega)}{m^{*}r_0}\right|^2\sum_{l,m}\int_{E_F}^{E_F+\hbar \omega} d\epsilon \frac{\epsilon(\epsilon - \hbar\omega)}{\hbar^4\omega^4} g_{l}^{sp} (\epsilon) \left[ g_{l+1}^{sp}(\epsilon - \hbar\omega)A_{lm}^2 + g_{l-1}^{sp}B_{lm}^2 \right] \\ & = \frac{4\pi}{\hbar} \left| \frac{eE(\omega)}{3 m^* r_0 \omega^2 \sqrt{\tilde{\epsilon}}}  \right|^2  \sum_l \int_{E_F}^{E_F+\hbar \omega} d\epsilon \sqrt{\epsilon(\epsilon-\hbar\omega)}\Theta(\epsilon-E_l^\text{min})\left[ (l+1)\Theta(\epsilon-\hbar\omega - E_{l+1}^{\text{min}}) + l \Theta(\epsilon-\hbar\omega - E_{l-1}^{\text{min}})\right ],
\end{align}
\end{widetext}
where the angular momentum selection rules were used and the resulting summation over $m$ was carried out. The integral over $\epsilon$ can be performed assuming that $l'=l\pm1 \approx l$ (which is valid in the limit of large radii). The resulting sp-to-sp hot carrier rate is expressed as a sum of several terms according to 
\begin{widetext}
\begin{equation}
    N^{sp \rightarrow sp}(\omega) = C(\omega) [ N_a(\omega) + N_b(\omega,l_{F}) - N_b(\omega,l_{0}) + N_{c}(\omega,l_{F}) - N_{c}(\omega,l_{0}) + N_{d}(\omega) ]
\end{equation}
with
\begin{align}
\label{eq: Na}
& N_{a}(\omega) = \left[ \sqrt{E_F(E_F + \hbar\omega)} (2E_{F} + \hbar\omega) - \sqrt{E_F(E_F - \hbar\omega)} (2E_{F} - \hbar\omega) + \hbar^2\omega^{2}\ln{\frac{\sqrt{E_{F}} + \sqrt{E_{F} - \hbar\omega}}{\sqrt{E_{F}} + \sqrt{E_{F} + \hbar\omega}}}\right](1+l_{0})^{2}, \\
\label{eq: Nb}
& N_{b}(\omega,l) = \frac{\alpha(l)}{120 \tilde{\epsilon}}\left( 24\hbar^{2}\omega^{2} + (2+l)^{2} \left[(-2 + 20l)\omega \tilde{\epsilon} + (l+2)^{2}(5l+1)\tilde{\epsilon}^{2} \right]\right),\\
\label{eq: Nc}
& N_{c}(\omega,l) = 
\left(\frac{\hbar^{3}\omega^{3}}{\tilde{\epsilon}} -\frac{9\hbar^2\omega^{2}}{8} \right) \ln \left[{\frac{\alpha(l) + \tilde{\epsilon}(l+2)}{{\alpha(l) - \tilde{\epsilon}(l+2)}}}\right] - \frac{\hbar^2\omega^{2}}{4}(2l+1)^{2} \ln \left[{\frac{\alpha(l) + \tilde{\epsilon}(l+2)}{{2\sqrt{\tilde{\epsilon}}~(\sqrt{E_{F}} + \sqrt{E_{F} + \omega})}}}\right] + \frac{\hbar^{2}\omega^{2}(4-l)\alpha(l)}{2\tilde{\epsilon}},\\
\label{eq: Nd}
& N_{d}(\omega) = \sqrt{E_F(E_F + \hbar\omega)} (2E_{F} + \hbar\omega)((l_{F}+1)^{2}-(l_{\rm 0})^{2}),
\end{align}
where we defined $C(\omega) =2|eE(\omega)|^{2}/(3m^{*}\pi\hbar^{3}\omega^{4})$, $\alpha(l) = 2\tilde{\epsilon} \sqrt{(l/2+1)^{2} + \omega/\tilde{\epsilon}}$, $l_{\rm 0} = 2(\sqrt{(E_{F}-\hbar\omega)/\tilde{\epsilon}}-1)$  and $l_{F} = 2(\sqrt{E_{F}/\tilde{\epsilon}}-1)$.
\end{widetext}

%\begin{widetext}
%\begin{align}
%N_{b1}^{s \rightarrow s}(\omega) &= -C(\omega)\frac{2 \omega^{3} \ln[{\alpha(l) +  \tilde{\epsilon}(l+2)}]}{\tilde{\epsilon}} + \frac{\alpha(l)\omega^{2}}{120 \tilde{\epsilon}} \left[ 96-60l + \frac{\tilde{\epsilon}}{\omega}(l+2)^{2}(5l+1)(\frac{\tilde{\epsilon}}{\omega}(l+2)^{2} - 2) \right]\\
%N_{b2}^{s \rightarrow s}(\omega) &= C(\omega)\omega^{2}\left((l^{2} +l-2) \ln \left[{\frac{\alpha(l) + \tilde{\epsilon}(l+2)}{2\sqrt{\tilde{\epsilon}}}}\right] - \frac{(l-4)\alpha(l)}{2\tilde{\epsilon}} + \frac{2\omega}{\tilde{\epsilon}} \ln \left[{\frac{\alpha(l) + \tilde{\epsilon}(l+2)}{\sqrt{\tilde{\epsilon}}}}\right] \right)\\
%N_{b3}^{s \rightarrow s}(\omega) &= C(\omega)\left( \sqrt{E_F(E_F + \omega)} (2E_{F} + \omega) - \omega^{2} \ln\left[{\sqrt{E_{F}} + \sqrt{E_{F} + \omega}}\right]\right) \sum^{l_F}_{l_{\rm 0}} (2l+1)
%\end{align}
%\end{widetext}

\subsection{Material parameters}

To evaluate the above expressions for the hot carrier generation rates, the values of several material-specific parameters are required. Here, we focus on silver nanoparticles. In the Drude model for the bulk dielectric constant, we use $\hbar\gamma_P=60$~meV~\cite{Sonnichsen2002}, $\hbar\omega_0=9.07$~eV and $\epsilon_b=4.18$~ \cite{Manjavacas2014} with a field intensity $E_{0} = 8.68 \cdot 10^{5}$~V/m. The Fermi energy is set to $E_F=5$~eV and $\epsilon_d=2$~eV~\cite{Cazalilla2000}. For simplicity the effective mass is set to the bare electron mass. 

%JL: changed dipole moment in response to reviewer 1
To determine the atomic transition dipole moment $z^\text{at}_{dp}$, we carry out a first-principles density-functional theory (DFT) calculation for an isolated silver atom. We use the all-electron code FHI-aims~\cite{Blum2009}, the PBE exchange-correlation functional~\cite{Perdew1996} and the default "tight" atom-centered numerical basis sets~\cite{Blum2009}. We average the square of the transition dipole moments of all possible d-state to p-state transitions and take the square root of this value. This results in $z^\text{at}_{dp}/e=$0.13~\AA.

We also carry out a first-principles DFT calculation for a fcc silver crystal (with $a_0=4.145$~\AA) using the same parameters as for the silver atom. A $12 \times 12 \times 12$ k-point grid was used to sample the first Brillouin zone. From a Mulliken analysis of the sp-band we determine its average p-state content in the k-space region where the sp-band lies above the Fermi level (as this region is of relevance to d-to-s transitions). This procedure yields $\alpha_p=0.81$.

\begin{figure}
    \centering
    \includegraphics[width =0.45\textwidth]{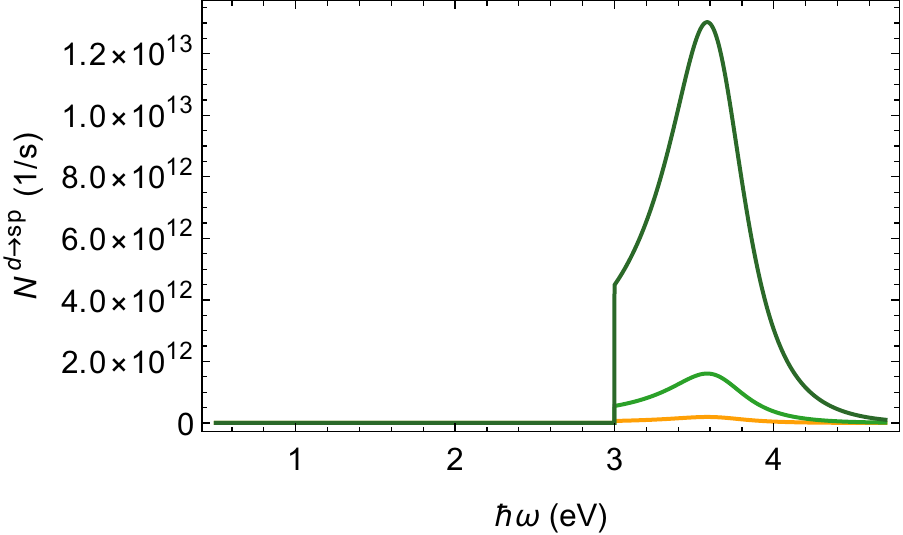}
    \caption{Hot carrier generation rate due to d-to-sp transitions as a function of photon energy for silver nanoparticles with different radii. We present results for $r_0=5$~nm (yellow curve), $r_0=10$~nm (light green curve) and $r_0=20$~nm (dark green curve).}
    \label{fig: Nds}
\end{figure}

\begin{figure}
    \centering
    \includegraphics[width=0.45\textwidth]{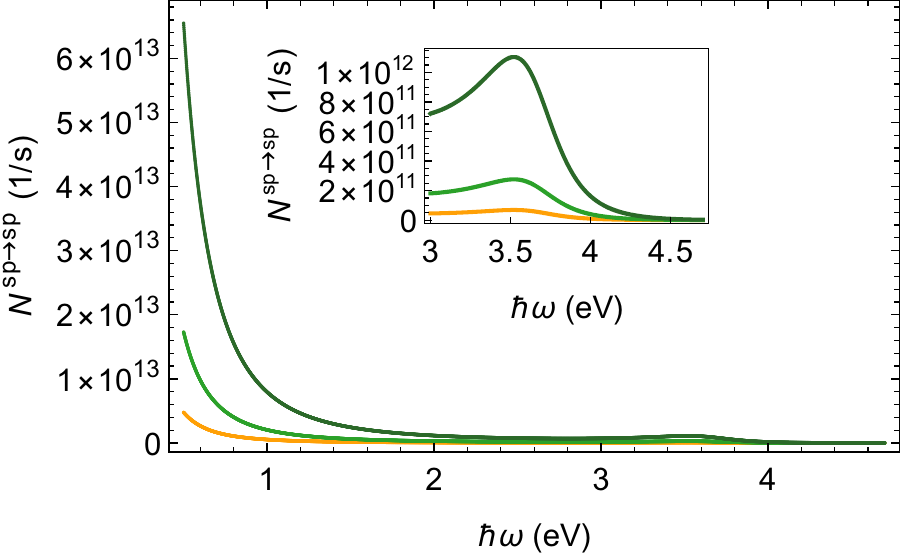}
    \caption{Hot carrier generation rate due to sp-to-sp transitions as a function of photon energy for silver nanoparticles with different radii. We present results for $r_0=5$~nm (yellow curve), $r_0=10$~nm (light green curve) and $r_0=20$~nm (dark green curve).}
    \label{fig: Nss}
\end{figure}

\begin{figure}
    \centering
    \includegraphics[width=0.45\textwidth]{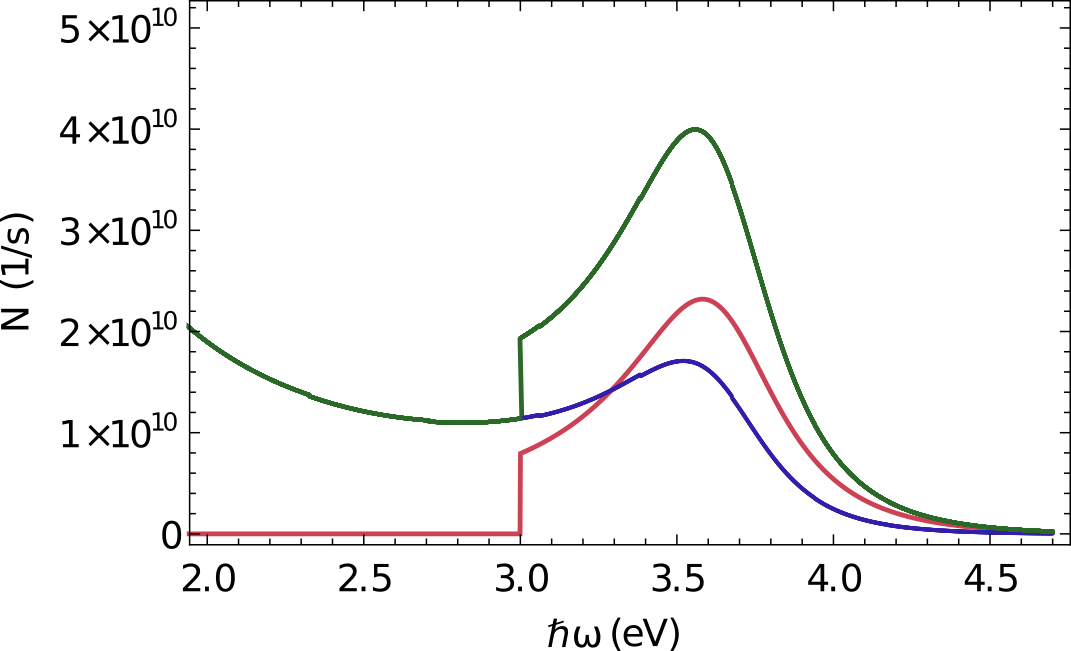}
    \caption{Plasmonic hot carriers generation rate due to d-to-sp transitions (red curve), sp-to-sp transitions (blue curve) and total generation rate (green curve) as a function of photon energy for a silver nanoparticle with $r_0=2.5$~nm.}
    \label{fig: NsdVsNss}
\end{figure}

\begin{figure}
    \centering
    \includegraphics[width=0.45\textwidth]{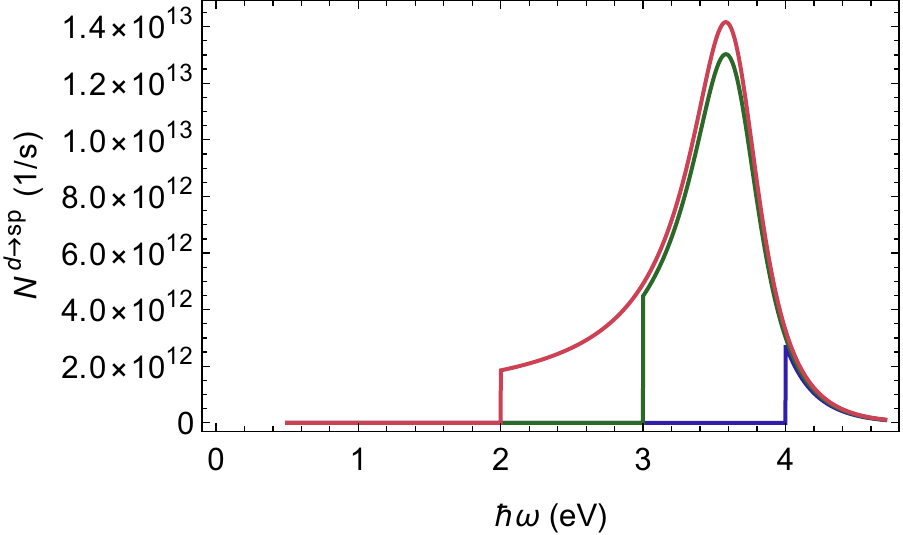}
    \caption{Hot carrier generation rate due to d-to-sp transitions as a function of photon energy for silver nanoparticles ($r_0=20$~nm) for $\epsilon_{d} = 1$~eV (blue curve), $\epsilon_{d} = 2$~eV (green curve) and $\epsilon_{d} = 3$~eV (red curve).}
    \label{fig: NdEd}
\end{figure}

\section{Results}
%LRC: Here I have added the comment about the r0^3 dependence
%JL: done above already
Figure~\ref{fig: Nds} shows the contribution of d-to-sp transitions to the total hot carrier generation rate in Ag nanoparticles with different radii as function of photon energy. No transitions occur when the photon energy is smaller than the energy difference between $E_F$ and $\epsilon_d$. The rate exhibits a peak at the localized surface plasmon energy, $\hbar \omega_{LSP}=3.65$~eV, as a result of the strong LSP-induced enhancement of the electric field. As discussed above, the rate scales with the volume of the nanoparticle, see Eq.~\eqref{eq:Nds_large}, explaining the rapid increase of the d-to-sp excitation rate with increasing nanoparticle size.

Figure~\ref{fig: Nss} shows the contribution of sp-to-sp transitions. For small photon energies, the hot carrier rate diverges. This is a consequence of the large matrix elements for transitions between states that are close to the Fermi energy, see Eq.~\eqref{eq:yannouleas}, which is also consistent with the trends observed in time-dependent DFT calculations \cite{RomanCastellanos2019}. Similarly to the d-to-sp case, the curves exhibit a peak when the photon energy is sufficiently large to excite a localized surface plasmon, see inset of Fig.~\ref{fig: Nss}.

In contrast to d-to-sp interband transitions, vertical sp-to-sp transitions are not possible in the bulk material. However, such transitions are found in nanostructures where translational invariance is broken and crystal momentum is not conserved due to the presence of the nanoparticle surface. As a consequence, we expect the sp-to-sp transition rate to scale with $r_0^2$, i.e. with the area of the nanoparticle surface. Comparing Fig.~\ref{fig: Nds} and Fig.~\ref{fig: Nss}, we indeed find that d-to-sp transitions dominate, but that sp-to-sp transitions become increasingly important as the size of the nanoparticle is reduced (of course, sp-to-sp transitions are always relevant at low photon energies because of their large matrix element, but those transitions are not relevant for the plasmon decay). The crossover to a size regime where sp-to-sp transitions dominate occurs at a radius of approximately 2-2.5~nm. Fig.~\ref{fig: NsdVsNss} shows the hot carrier rates for $r_0=2.5$~nm and it can be seen that the sp-to-sp and d-to-sp contributions are of comparable magnitude for this nanoparticle size.

Finally, we explore the dependence of the d-to-sp hot carrier rate on the energy of the d-band $\epsilon_d$. Fig.~\ref{fig: NdEd} shows results for $\epsilon_d=3$~eV and $\epsilon_d=1$~eV in addition to $\epsilon_d=2$~eV. For the lower d-band energy, the onset of d-to-sp transitions occurs at energies larger than the LSP energy. As a consequence, the d-to-sp rate does not "benefit" from the LSP electric field enhancement and the overall d-to-s contribution is relatively small. 

%LRC: Added sentence about first principles calculations and other decay channels.
\section{Conclusions}
We have presented a new approach for calculating the contribution of d-bands to the hot carrier generation rate in metallic nanoparticles. In particular, we have derived an equation for the envelope function of a nanoparticle state that derives from a d-band and solved this equation in the limit of perfectly flat bands. Evaluating Fermi's golden rule for d-band state to sp-band state transitions and also for sp-band state to sp-band state transitions in silver nanoparticles, we find that the rate of d-to-sp transitions scales with the nanoparticle volume and dominates hot carrier generation for nanoparticles with radii larger than 2.5~nm. In contrast, the rate of sp-to-sp transitions scales with the surface area of the nanoparticle and gives the dominant contribution for very small systems. It should be straightforward to extend this approach to other materials, such silver and gold alloys, and their nanostructures 
allowing to refine the design of nanoplasmonic devices with optimized hot carrier rates. Future work will be carried out to compare our predictions to first-principles calculations and to extend the description to other loss mechanisms, such as phonon-mediated processes.

\section{Acknowledgements}
The authors acknowledge support from the Thomas Young Centre under grant no. TYC-101. This work was supported through a studentship in the Centre for Doctoral Training on Theory and Simulation of Materials at Imperial College London funded by the EPSRC (EP/L015579/1) and  through EPSRC projects EP/L024926/1 and EP/L027151/1. J.M.K. and J.L. acknowledge support from EPRSC under Grant No. EP/R002010/1. Via J.L.'s membership of the UK's HEC Materials Chemistry Consortium, which is funded by EPSRC (EP/L000202), this work used the ARCHER UK National Supercomputing Service.

\bibliography{libraryCLEAN}
\end{document}